\documentclass[aps, prl, amsmath, floats, floatfix, twocolumn, nofootinbib, superscriptaddress, showpacs]{revtex4-1}
\usepackage{graphicx}
\usepackage{color}
\usepackage{latexsym}
\usepackage{amsfonts}
\usepackage[colorlinks=True]{hyperref}

\newcommand{\be}{\begin{eqnarray}}
\newcommand{\ee}{\end{eqnarray}}

\begin{document}

\title{Theory-agnostic tests of gravity with black hole shadows}

\author{Sourabh~Nampalliwar}
\email[Corresponding author: ]{sourabh.nampalliwar@uni-tuebingen.de}
\affiliation{Theoretical Astrophysics, Eberhard-Karls Universit\"at T\"ubingen, 72076 T\"ubingen, Germany}

\author{Saurabh~K.}
\affiliation{PDPIAS, Charusat University, Anand, Gujarat 388421, India}

\date{\today}

\begin{abstract}
Observations of black hole shadows with the Event Horizon Telescope have paved way for a novel approach to testing Einstein's theory of general relativity. Early analyses of the measured shadow put constraints on theory-agnostic parameters typically used to study deviations from Einstein's theory, but the robustness of these constraints was called into question. In this letter, we use a generic theory-agnostic metric to study the robustness of parameter estimation with BH shadows, taking into consideration current measurements made with the Event Horizon Telescope and future measurements expected with the Event Horizon Imager. We find that the robustness issue is highly nuanced, and parameter constraints can be highly misleading if parameter degeneracy is not handled carefully. We find that a certain kind of deviation is particularly well suited for the shadow based analysis, and can be recovered robustly with shadow measurements in the future.
\end{abstract}

\maketitle

\textit{Context} $-$ General relativity serves as the standard framework for describing gravitational phenomena in our universe. While extremely precise in the so called \textit{weak-field} regime, tests in the \textit{strong-field} regime are getting progressively better, with the detection of gravitational waves~\cite{LIGOScientific:2019fpa,Abbott:2020jks}, x-rays~\cite{Cao2017,Abdikamalov:2019zfz}, and, most recently, black hole imaging~\cite{EventHorizonTelescope:2019ggy}. The horizon-scale imaging of the central region of the M87 galaxy using the Event Horizon Telescope heralds a new avenue for strong-field tests of gravity~\cite{PhysRevLett.125.141104,Volkel:2020xlc,Psaltis:2020ctj}, with upcoming observations of the center of the Milky Way promising to be a significantly step forward in this direction. All these tests exploit the prediction of general relativity (GR) that black holes (BHs) exist in our universe. Being highly compact, BHs reveal the nature of gravity in regions that are, simultaneously, highly relativistic and strongly curved. Starting out as mathematical idealizations, BHs are now thought to be extremely commonplace in our universe, coming in all sizes (from $\sim3M_{\odot}$ to $\sim10^{11}M_{\odot}$, where $M_{\odot}$ is the mass of the sun) and described by very few parameters (typically, only their mass and spin) within GR. All these properties make BHs the ideal systems for performing strong-field tests of gravity.

Although GR is the leading theory of gravity at present, it suffers from challenges on the theoretical side (e.g., incompatibility with quantum mechanics) as well as the observational side (e.g., dark matter and dark energy), which have led to proposals which modify/extend it. One approach, then, to test gravity is the start with one such modification/extension, make predictions, and test those predictions (as done in, for instance, Ref.~\cite{Mizuno:2018lxz}). This is extremely difficult to do in practice. A second approach has instead gained popularity, where one performs parametric modifications to the BH solution of GR and tests, against observations, whether the non-GR parameters show any deviation from their GR values. A violation of GR can be captured with such an approach and, in some cases, the non-zero non-GR parameters of such a theory-agnostic approach can be mapped to several specific theories and thus provide constraints on all of them simultaneously.

A series of recent works have looked at the possibility of using BH shadow measurements by the Event Horizon Telescope (EHT) to perform theory-agnostic tests. The EHT collaboration has, in Ref.~\cite{PhysRevLett.125.141104}, put constraints on the leading-order deviation parameters of some parametrized metrics, under the assumption that higher-order deviation parameters will have a progressively weaker effect on the shadow. This assumption was challenged in Ref.~\cite{Volkel:2020xlc}, where the authors show that higher-order parameters can completely wash out the constraints on the leading-order deviation parameters. Similar washout was observed in Ref.~\cite{Psaltis:2020ctj} for the first- and second-order Post-Newtonian parameters. Comparison of these constraints with those from other techniques have also been made~\cite{Nampalliwar:2019iti,Psaltis:2020ctj,Glampedakis:2021oie}. The picture that emerges from all this suggests that, even if they can be observed and measured with some precision, BH shadows cannot provide meaningful constraints on parametrized deviations away from the Kerr metric. In this letter, we revisit this issue in more generic settings by $1)$ considering spinning BHs and associated non-GR deformations, $2)$ use measurements related to the size as well as the shape of the shadow, $3)$ use current as well as \textit{future} estimates of the BH shadows. We find that, in agreement with previous works, parametrized deviations suffer from degeneracies that make their estimation difficult. However, for certain kinds of parametric deviations, we find that BH shadows do provide good constraints, and these constraints are indeed robust. 

Since different observational techniques probe different aspects of the spacetime, robust constraints obtained from BH shadows provide an independent and unique estimate, which can be compared and combined with the constraints from other techniques, to improve the constraints as well as break inter-parameter degeneracies. As several techniques to perform strong-field tests of gravity emerge and improve in the coming years, this kind of synergy will become extremely valuable.  

Throughout the letter, we work in the units of $G=c=M=1$.

\textit{Methods} $-$ 
While colloquially called the BH \textit{shadow}, the observable that is typically used in such analyses is the \textit{apparent boundary}, which is the gravitationally lensed image of the photon orbit~\cite{bardeen1973}. The photon orbit marks the location where photons, under the extreme gravity of the BH, travel in orbits of constant radius. For spherically symmetric BHs, these quantities can be written analytically as
\be
	r_{\textrm{ph}} = \left (\left . \frac{\textrm{d}}{\textrm{d}r}\ln\sqrt{-g_{tt}}\right )^{-1} \right |_{r_{\textrm{ph}}} 
\ee 
and 
\be
	r_{\textrm{sh}} = \frac{r_{\textrm{ph}}}{\sqrt{-g_{tt}(r_{\textrm{ph}})}},
\ee
where $r_{\textrm{ph}}$ is the photon orbit, $r_{\textrm{sh}}$ is the apparent boundary and $g_{tt}$ is the $tt$-component of the metric. In particular, for a Schwarzschild BH, $r_{\textrm{ph}}=3$ and $r_{\textrm{sh}}=3\sqrt{3}$~\cite{Luminet:1979nyg}. For generic spinning BHs, one instead needs to ray-trace photons and compute the apparent boundary numerically. We use the numerical scheme described in Ref.~\cite{Nampalliwar:2020asd} and based on Ref.~\cite{relxillnk}, placing the observer at the equatorial plane and at some large distance where the spacetime is effectively flat. We ray-trace 30 photons in this manner to create a discrete apparent boundary (ray-tracing more photons has a negligible effect on the observables). We take the average shadow size, defined as
\be
	\tilde{D} = \frac{1}{N}\sum_{i=0}^{N} D_i,
\ee
where $D_i$ is the shadow diameter along a specific orientation, as our first observable. For a Schwarzschild BH, $D_i$'s, as well as $\tilde{D}$, are equal to $2r_{\textrm{sh}}$.


While for a spherically symmetric BH metric, the shadow is perfectly circular, BH spin and deviation parameters introduce \textit{non-circularity} in the shadow~\cite{Johannsen:2010ru}. This non-circularity is strongly suppressed for a source like M87$^*$ whose spin axis is nearly aligned with its line of sight to Earth. However, for a nearly equatorial line of sight, as could be the case for Sgr~A$^*$~\cite{Li:2013xka,Psaltis:2014dea}, non-circularity might be as important as the size of the shadow in estimating the parameters of the BH. The EHT collaboration used the standard deviation of the normalized shadow diameter, i.e., $D_i/\tilde{D}$, along different orientations as a measure of the shadow's deviation from circularity~\cite{EventHorizonTelescope:2019ggy}. (To be precise, what is measured is the standard deviation of the image diameter divided by the mean diameter, and the image diameter and the shadow diameter are related by the scaling factor $\alpha$.) We use this quantity, calling it the shadow shape, as our second observable.

The EHT has been observing Sgr A$^*$ and is expected to release high-resolution images (or a movie) in the near future. In the absence of actual measurements, we will assume two \textit{potential} measurements of the shadow observables, possible in the near and the far future, of Sgr A$^*$. The first, which we term \textit{current} measurement, is based on the EHT measurement of M87$^*$. These are
\be\label{eq:current}
	\tilde{D} = 3\sqrt{3} \pm 17\%, \qquad \sigma_D \lesssim 0.05,
\ee
where $\sigma_D$ is the shadow shape observable discussed above. The second, which we term \textit{future} measurement, is expected to be within the capabilities of very long baseline interferometry in the coming decades. These are
\be\label{eq:future}
	\tilde{D} = 3\sqrt{3} \pm 5\%, \qquad \sigma_D \lesssim 0.01.
\ee
This is achievable, for instance, with the space-based Event Horizon Imager interferometer array~\cite{Roelofs:2021wdi}, which is currently in the planning stage. 

\textit{Deviation parameters} $-$ There are several parametrically deformed Kerr metrics available in literature. These metrics distinguish themselves in the symmetries they carry over from the Kerr metric (e.g., separability of Hamilton-Jacobi equations, separability of Klein-Gordon equations) and the parametrization they use (e.g., power series, continued fractions). One of the best metrics in this context in the one proposed in Ref.~\cite{Konoplya2016}. It is axisymmetric and asymptotic flat and employs a continued fraction expansion of the parametric deviation functions, allowing for rapid convergence. In Boyer-Lindquist-like coordinates, the metric can be written as
\be\label{eq:metric}
ds^2 &=& - \frac{N^2 - W^2 \sin^2\theta}{K^2} \, dt^2 - 2 W r \sin^2\theta \, dt \, d\phi
\nonumber\\ &&
+ K^2 r^2 \sin^2\theta \, d\phi^2 
+ \frac{\Sigma \, B^2}{N^2} \, dr^2 + \Sigma \, r^2 \, d\theta^2 \, ,
\ee
where $\Sigma = 1 + a_*^2\cos^2{\theta}/r^2$ and $a_*$ is the dimensionless spin parameter. $N^2$, $W$, $K^2$ and $B^2$ encode deviations away from the Kerr metric and are functions of $r$ and $\theta$. Their expansion, and their forms in the Kerr limit, can be found in Ref.~\cite{Nampalliwar:2019iti}. Here we are interested in two of these deviation functions, viz. $N^2$ and $W$. $N^2$ primarily affects the $tt$ and the $rr$ components of the metric, and $W$ primarily affects the $t\phi$, $tt$ and $\phi\phi$ (via $K^2$, see Eq. (23d) in Ref.~\cite{Konoplya2016}) components of the metric. These two functions tend to have the strongest impact on the shadow, and therefore are expected to be constrained to some degree with our observables. They can be written as (in terms of the $\cos{\theta}-$based series expansion defined in Eqs. (23) in Ref.~\cite{Konoplya2016}, we are neglecting terms of order $\cos{\theta}$ and higher)
\be
	N^2(r,\theta) = 1-\frac{2}{r} + \frac{a_*^2}{r^2} + \left(1-\frac{2}{r}\right)\frac{r_0^3}{r^3}\,\widetilde{A}_0(r)
\ee
and
\be
	W(r,\theta) = \frac{2\,a_*}{r^2 + a_*^2\cos^2{\theta}} + \frac{1}{\Sigma}\frac{r_0^3}{r^3}\,\widetilde{W}_0(r),
\ee
where $r_0$ is the horizon's radius. In both equations, the last term on the right hand side parametrizes deviations from the Kerr metric, while the remaining terms ensure the correct Kerr limit. (obtained by using Eq. (4)--(7) in Ref.~\cite{Nampalliwar:2019iti} and setting all the deviation parameters to zero.) Here,
\begin{equation}
	\widetilde{A}_0 (x) = \cfrac{a_{01}}{1+\cfrac{a_{02}x}{1+\cfrac{a_{03}x}{1+\cdots}}}, \quad \widetilde{W}_0 (x) = \cfrac{w_{01}}{1+\cfrac{w_{02}x}{1+\cfrac{w_{03}x}{1+\cdots}}},
\end{equation}
where $x = 1-r_0/r$. We analyze parameters up to the second leading order, i.e., up to $a_{02}$ for $N^2$ and $w_{02}$ for $W$.


To do the analysis, we use the Python library \texttt{Gallifray}~\cite{gallifrayweb}, recently developed by us to do model comparison and parameter estimation of BH images obtained with very long baseline interferometry (VLBI). Here we use the parameter estimation module of library to perform an MCMC analysis, and the plotting module to present the one and two-dimensional posterior distributions. We assume uniform priors for the BH parameters, bounded between $\pm 1$ in the case of $a_*$, $\pm 5$ in the case of $a_{01}$ and $w_{01}$, and $\pm 20$ in the case of $a_{02}$ and $w_{02}$. Increasing the range of priors does not affect the qualitative inferences drawn below. Note that some of these parameters have their respective theoretical bounds, to ensure spacetime does not have pathologies, on top of these priors. For instance, large negative values of $a_{01}$ ($\lesssim -1$) give rise to singularities outside the horizon. The shadow size observable $\tilde{D}$ is assumed to have a normal distribution with the mean at $3\sqrt{3}$ and the variance determined by the uncertainty in shadow size (see Eqs.~\ref{eq:current} and~\ref{eq:future}). The shadow shape observable $\sigma_D$ is assumed to have a uniform distribution bounded between $0$ and the maximum uncertainty in shadow shape, given in Eqs.~\ref{eq:current} and~\ref{eq:future}. 
 
\begin{table}[!htb]
\caption{\label{t-cases} Summary of different cases considered and the parameters allowed to vary in each case. A checkmark indicates a variable parameter and a dash indicates a frozen parameter.}

\centering
\begin{tabular}{lcccccc}
\hline\hline
\hline
Parameter $\rightarrow$\hspace{0.2cm} & $a_*$\hspace{0.2cm} & $a_{01}$\hspace{0.2cm} & $a_{02}$\hspace{0.2cm} & $w_{01}$\hspace{0.2cm} & $w_{02}$\hspace{0.2cm} \\
\hline
 Case I & \checkmark\hspace{0.2cm} & \checkmark\hspace{0.2cm} & $-$\hspace{0.2cm} & \checkmark\hspace{0.2cm} & $-$\hspace{0.2cm}\\
Case II & \checkmark\hspace{0.2cm} & \checkmark\hspace{0.2cm} & \checkmark\hspace{0.2cm} & $-$\hspace{0.2cm} & $-$\hspace{0.2cm}\\
 Case III & \checkmark\hspace{0.2cm} & $-$\hspace{0.2cm} & $-$\hspace{0.2cm} & \checkmark\hspace{0.2cm} & \checkmark\hspace{0.2cm} \\
\hline\hline
\end{tabular}
\end{table}

To analyze constraints and degeneracies on these parameters using the BH shadow, we allow different sets of parameters to vary at a time, while fixing others to zero. The analysis is categorized in three cases (summarized in Tab.~\ref{t-cases}):  

\textit{Case I} $-$ In the first case, we vary $a_*$, $a_{01}$ and $w_{01}$, i.e., spin and the leading order deviation parameters of the $N^2$ and $W$ functions respectively. This choice is motivated by two considerations: Firstly, n\"{a}ively, one might think that higher-order parameters will have lesser impact on the observables and, consequently, neglecting them will only marginally affect the constraints on the leading-order parameters. However, as we shall see, this is not the case. Secondly, other experimental techniques discussed above (gravitational waves, x-rays, etc.) make similar choices about these theory-agnostic deviation parameters and are, depending on the technique, justified to do so. 

\begin{figure}[!htb]
\centering
    \includegraphics[width=\columnwidth]{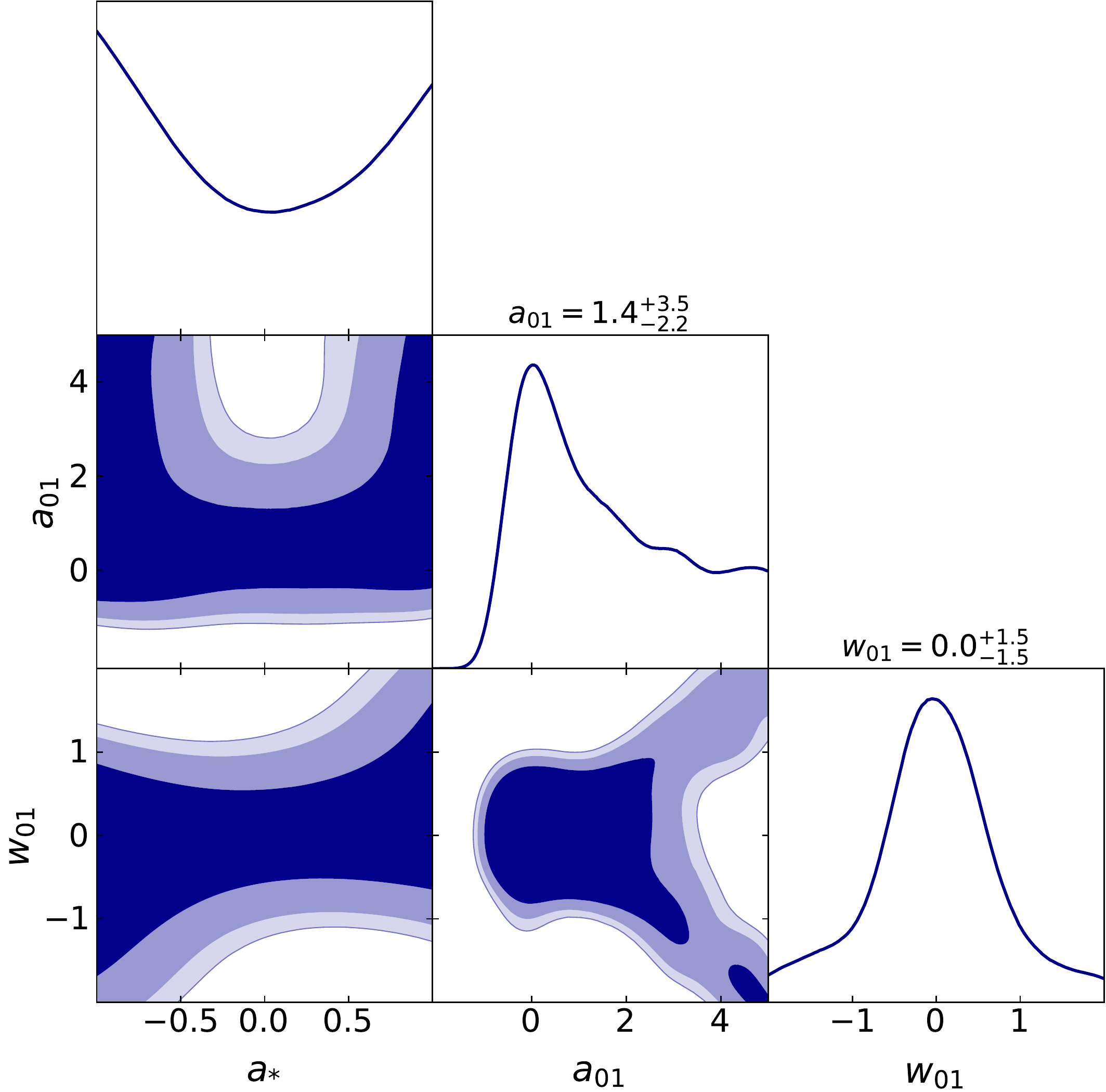}
    \includegraphics[width=\columnwidth]{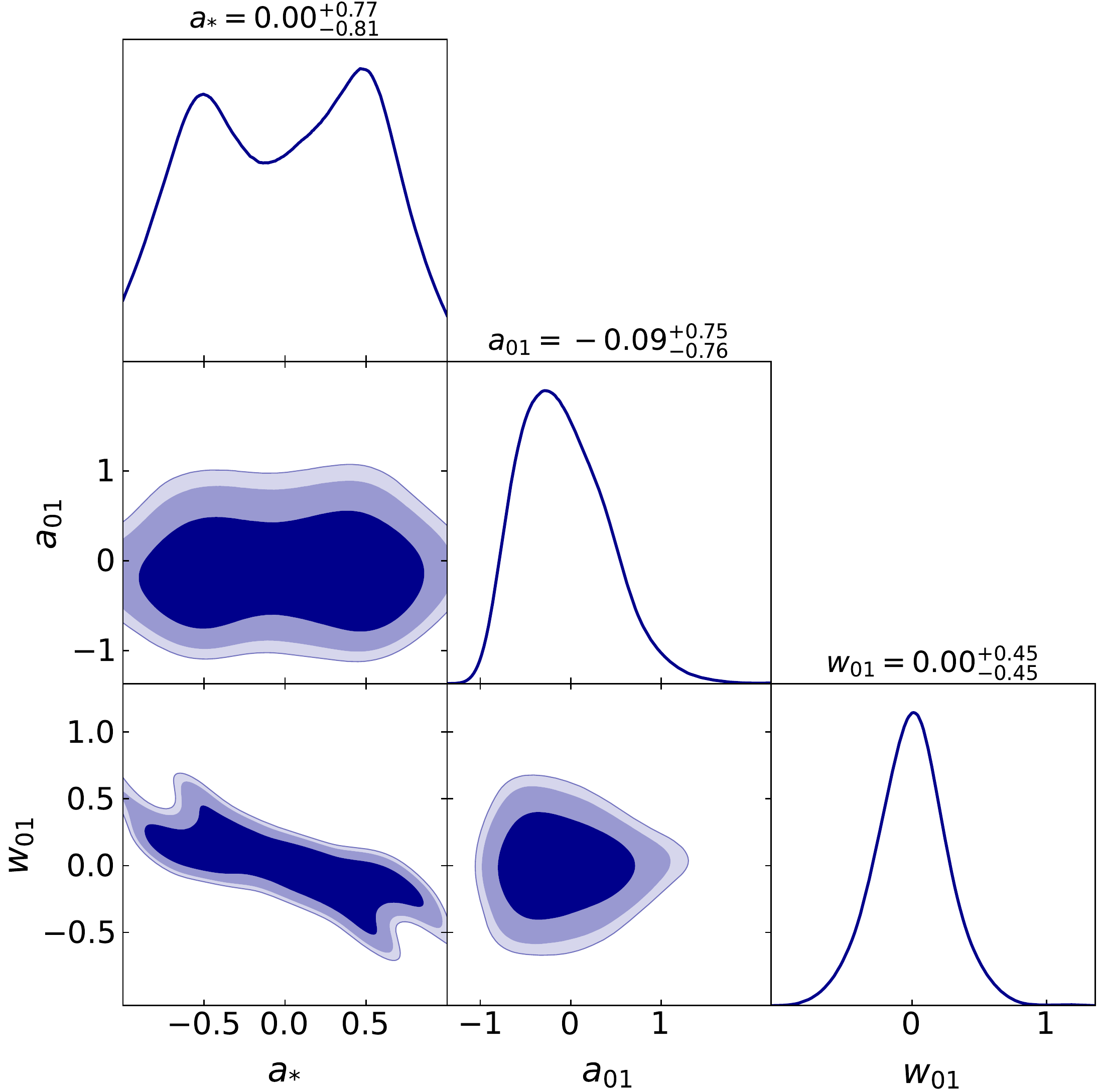}
    \caption{\label{fig:case1} Posterior distributions of the parameters analyzed in Case 1, viz., $a_*$, $a_{01}$ and $w_{01}$, with current (top panel) and future (bottom panel) measurements. Shown are the two-dimensional joint posteriors for each pair of parameters, along with contours at 68\%, 90\% and 99\% confidence levels, and one-dimensional posterior distribution for each parameter. See the text for more details.}
\end{figure}
The constraints on the three parameters with current measurements are presented in Fig.~\ref{fig:case1}. The figure shows the one-dimensional posterior distribution of each parameter along the diagonal panels and the two-dimensional posteriors along the off-diagonal panels. Several very interesting features become apparent in this figure. Starting with the current measurements, we see that although for zero spin there are some bounds on $a_{01}$ and $w_{01}$, free spin washes out both these bounds. Furthermore, assuming one deviation parameter to be zero, we see from the plot that the other has some bounds, but these also get washed out if both deviation parameters are allowed to vary. The posteriors shown here extend to the edges of the prior distribution, and this behavior continues for enlarged priors. In summary, with current measurements, no bounds are possible on the BH spin, $a_{01}$ and $w_{01}$ if all of them are set free. It is remarkable to note that even at the leading-order in deviation and with two shadow observables, inter-parameter degeneracies are impossible to break with current measurements.

Results of a similar analysis with future measurements is presented in Fig.~\ref{fig:case1}, and some very interesting new features emerge in this scenario. The one-dimensional posterior distributions shrink to a great extent for both $a_{01}$ and $w_{01}$, as expected for a more precise measurement, however, the BH spin remains unconstrained. Inter-parameter degeneracies also reduce significantly, and $a_{01}$ ($w_{01}$) remains well-constrained even for non-zero $w_{01}$ ($a_{01}$), as can be seen from the two-dimensional posterior distribution of $a_{01}$ and $w_{01}$. Two-dimensional posteriors involving spin show an interesting feature: the $a_{01}$ bounds are nearly independent of spin, whereas the $w_{01}$ bounds show a healthy dependence on spin, to the extent of excluding $w_{01}=0$ at large positive or negative spins. Interestingly, allowing both $a_{01}$  and $w_{01}$ to vary simultaneously does not wash out the bounds on them, whereas no bounds are possible on spin. In conclusion, based on this case, we can say that the leading order deviation parameters $a_{01}$ and $w_{01}$ can be measured with a future measurement of the BH shadow, even in the absence of an estimate of the BH spin.

While the results discussed above paint an interesting picture of parametrized tests of gravity with BH shadows, their robustness is not at all obvious. For example, is the assumption made above, that higher-order deviation parameters can be neglected, valid? We explore this robustness question now with the remaining two cases.

\textit{Case II} $-$ 
\begin{figure}[!htb]
\centering
    \includegraphics[width=\columnwidth]{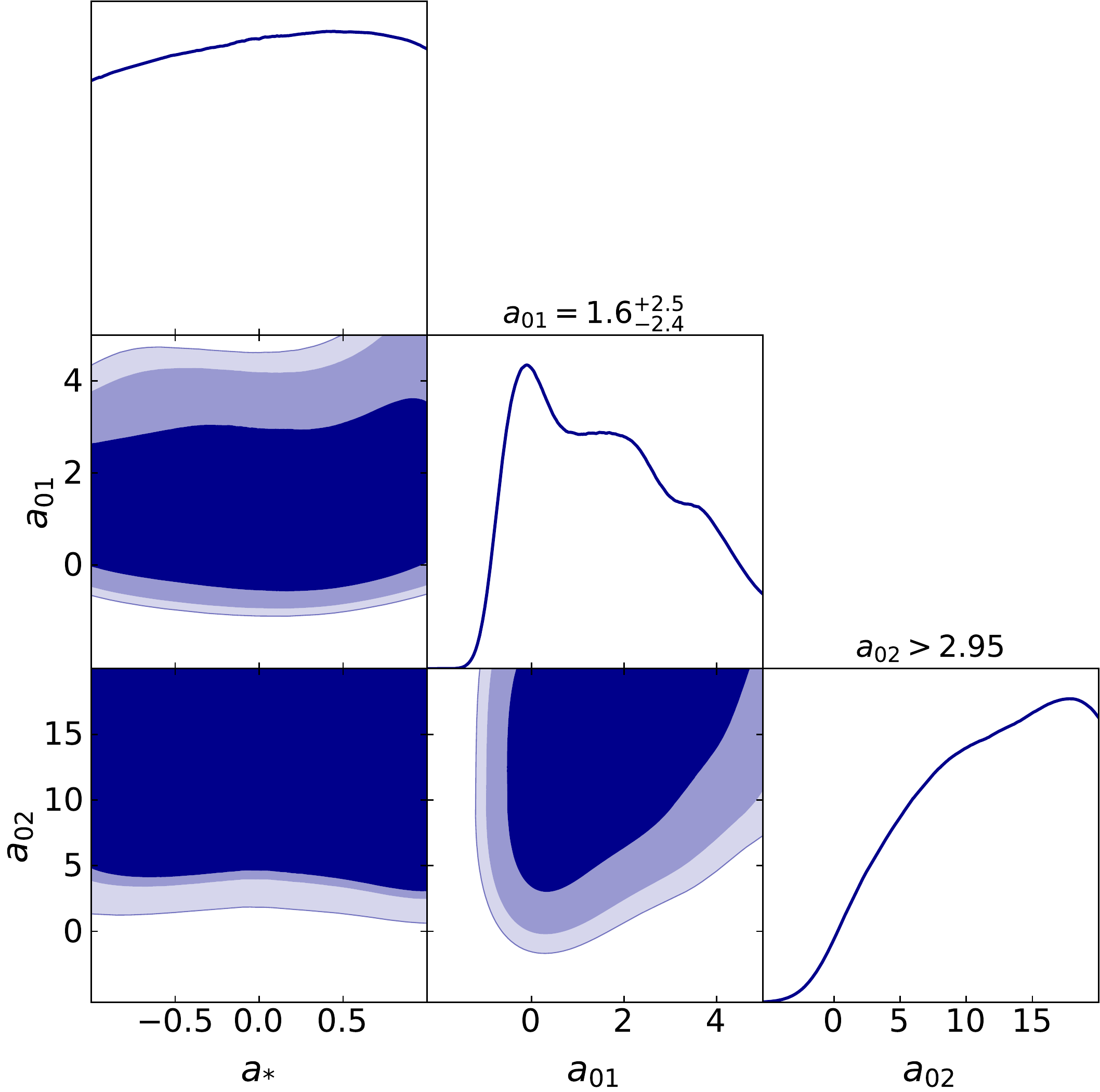}
%
    \includegraphics[width=\columnwidth]{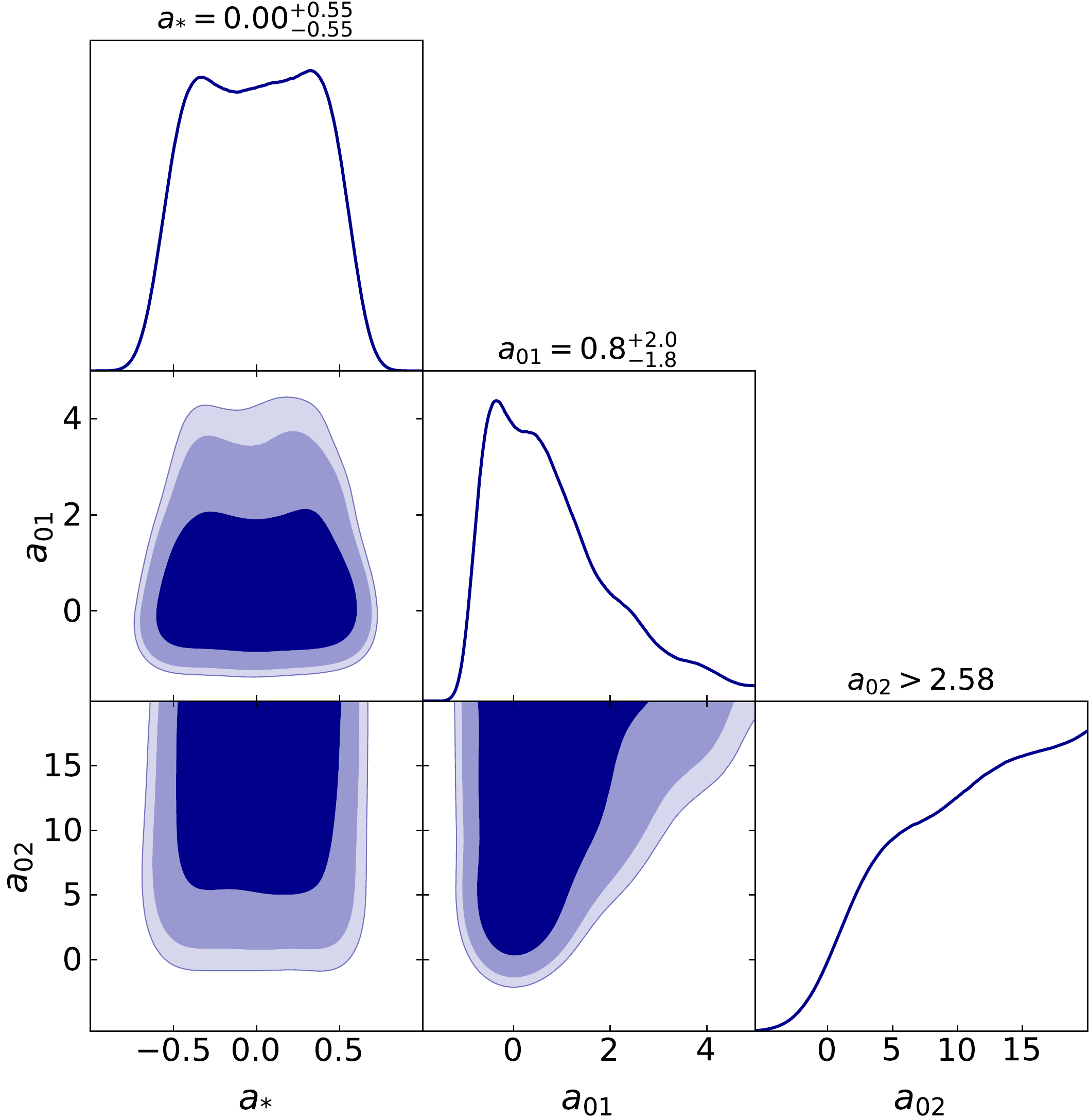}
    \caption{\label{fig:case2} Same as Fig.~\ref{fig:case1} for Case II.}
\end{figure}
We now vary $a_*$, $a_{01}$ and $a_{02}$, i.e., spin and the two lowest-order deviation parameters of the $N^2$ function, while setting all other deviation parameters to zero. The results are presented in Fig.~\ref{fig:case2}. $a_{02}$ is unbounded, extending all the way to the edge of the range of the prior, with both current and future measurements. $a_{01}$ also follows the trend, extending close to the edge of the range of the prior, and is strongly degenerate with $a_{02}$. Increasing the range of the $a_{02}$ prior makes the bounds on $a_{01}$ worse, and more precise measurements expected in the future do not seem to help. Thus, we find that the bounds obtained in Case I above on $a_{01}$ are not robust and get washed out in the presence of higher-order parameters. 

Interestingly, the constraints on BH spin follow the opposite trend. Spin is unbounded with current measurements but strongly bounded with future measurements. However, we have already seen in Case I that spin is unbounded with both measurements, so the bounds obtained on spin here are not robust either.

\textit{Case III} $-$
\begin{figure}[!htb]
\centering
    \includegraphics[width=\columnwidth]{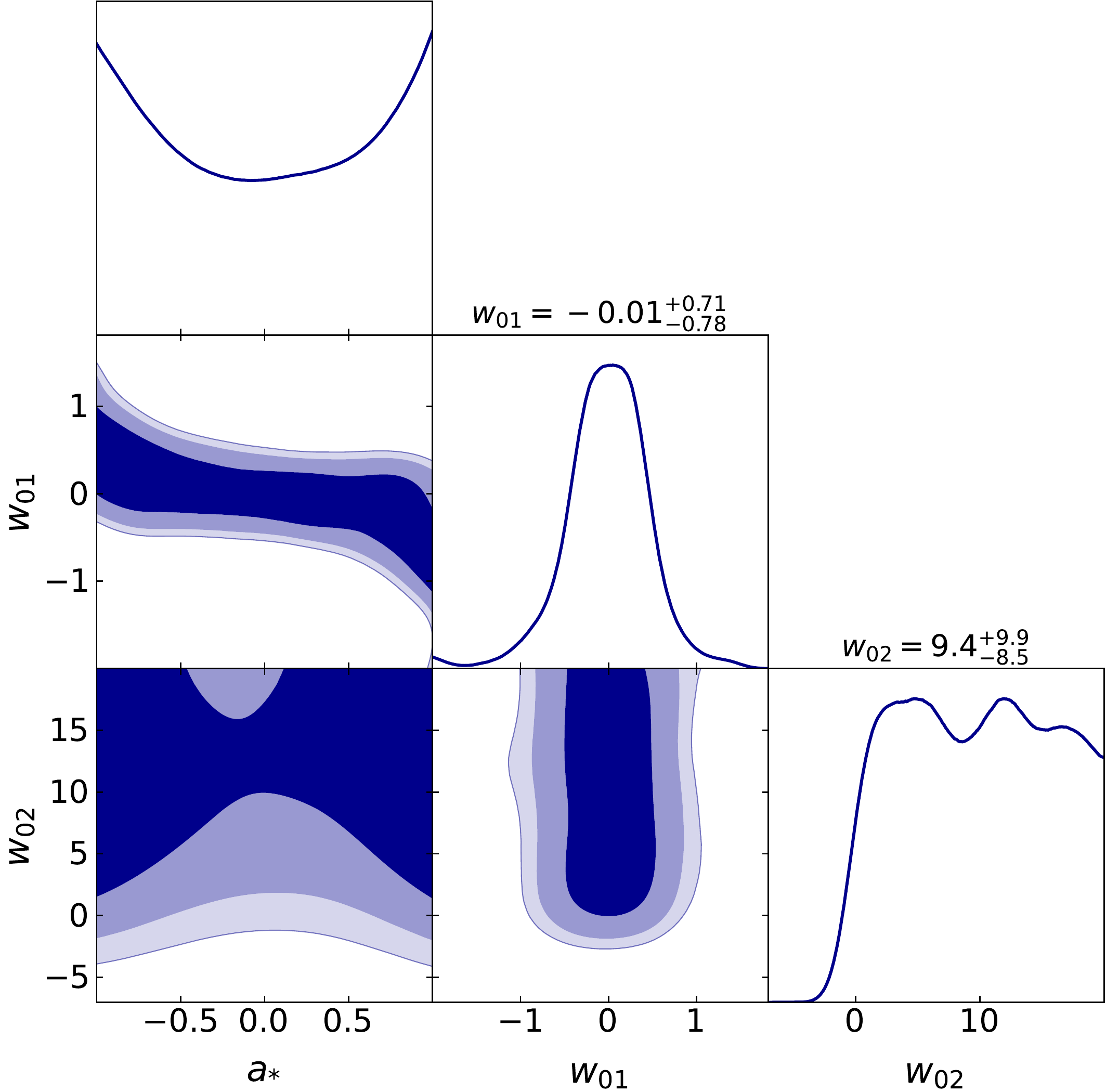}
%
    \includegraphics[width=\columnwidth]{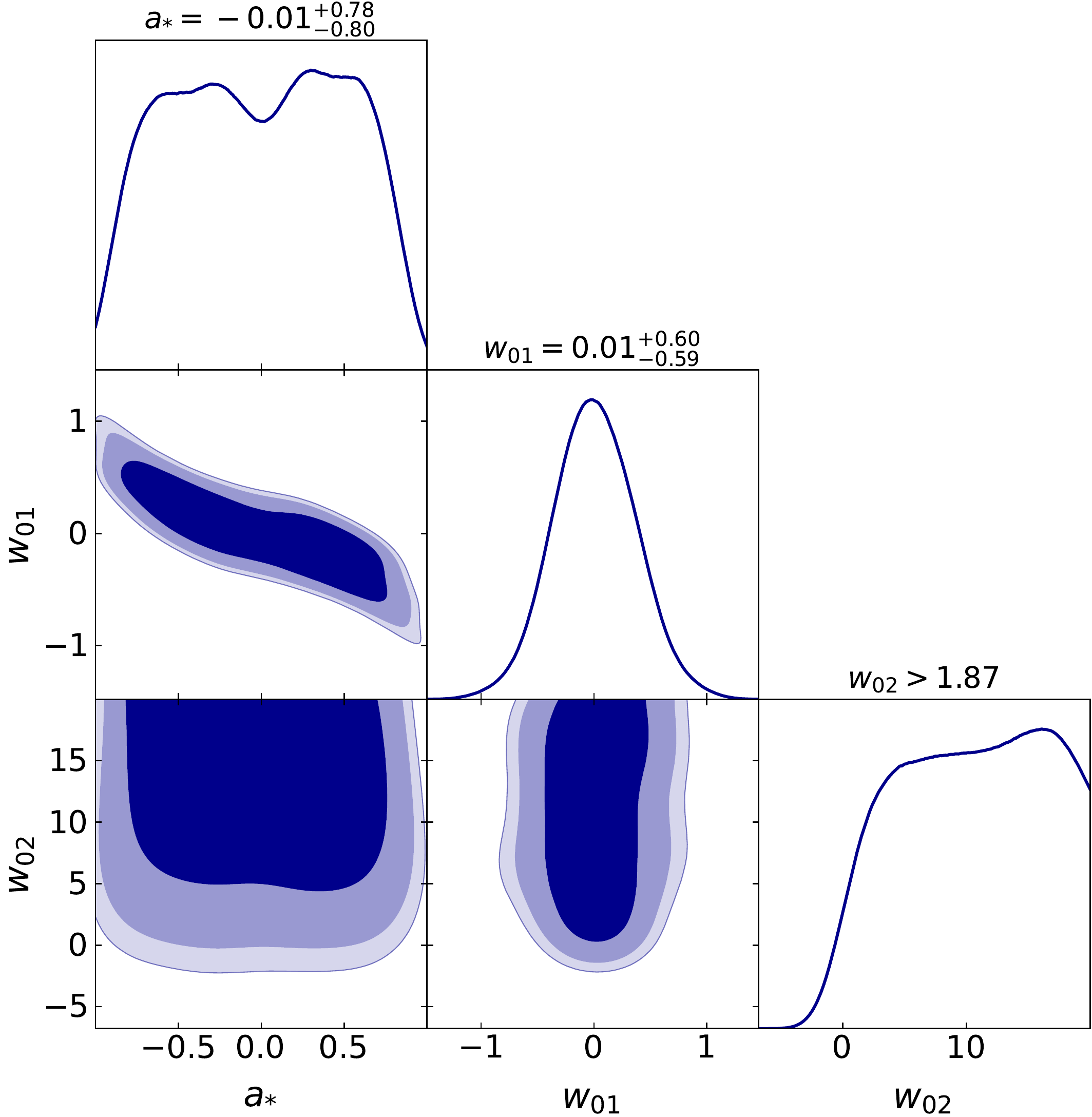}
    \caption{\label{fig:case3} Same as Fig.~\ref{fig:case1} for Case III.}
\end{figure}
We now come to Case III, where $a_*$, $w_{01}$ and $w_{02}$, i.e., spin and the two lowest-order deviation parameters of the $W$ function are free, while all other deviation parameters are set to zero. The results are plotted in Fig.~\ref{fig:case3} and we see a surprising feature: the bounds on $w_{01}$ are nearly independent of $w_{02}$. The origin of this behavior lies in the fact $w_{02}$ has an extremely weak effect on the observables, and does not grow with $w_{02}$. Furthermore, although the degeneracy with spin does affect it a little, $w_{01}$ has clear bounds both with current and future measurements. From Case I, we already know that the bounds on $w_{01}$ from current measurements are not robust, but $w_{01}$ can be constrained with future measurements even in the presence of $a_{01}$. Further analyses, allowing for simultaneous freedom of $a_*$, $a_{01}$, $a_{02}$, $w_{01}$ and $w_{02}$ leads to the same conclusion: $w_{01}$ can be constrained with future measurements \textit{in a robust way}.

Figs.~\ref{fig:case1}-\ref{fig:case3} also provide parameter estimates within 90\% uncertainty, however, we did not perform an exhaustive statistical analysis here. The primary result of this work is the qualitative idea that it is possible to constrain some deviation parameters with BH shadows in a robust way. There are some caveats though: the EHT array does not measure the shadow directly, rather, it measures a crescent whose size is proportional to the shadow size and the proportionality constant depends on the BH spin and possibly the deviation parameters. The crescent itself is a geometric approximation to a complex emission process happening in the BH vicinity, whose modeling with GRMHD simulations could add additional complexity to the estimation of parameters considered here. The assumption of an equatorial observer may not be valid. Finally, the future measurement considered here is based on anticipated technological progress and centered around a fiducial Schwarzschild BH (which would have $\tilde{D} = 3\sqrt{3}$ and $\sigma_D=0$); a measurement of similar precision but centered at different values may change the inferences drawn above.

{\it Conclusions} $-$
Parametrized test of gravity is currently a popular approach to study strong-field gravity, due in part to the complexity associated with analyzing alternative theories of gravity and to the plethora of such theories with no consensus on a preferred alternative. While tests with parametrized metrics alleviate these problems, they suffer from the problem of plenty -- with so many independent deviation parameters, measuring one robustly becomes nearly impossible. The problem is compounded by the fact that the parametrization is often based on an $M/r$ expansion which suffers from convergence issues close to the BH, precisely where the shadow originates from. These issues, if unresolved, can be serious enough to render parametrized tests of gravity with BH shadows impossible. In this letter, we show that the answer to the robustness question is highly nuanced. Ignoring higher-order parameters can lead to constraints that do not hold, as seen for $a_{01}$ in Cases I and II. Ignoring leading order parameters of different deviation functions can again lead to nonrobust constraints, as seen for BH spin in Cases I and II. At the same time, all is not lost -- we find that $w_{01}$ can indeed be constrained, with future measurements, in a robust way. It remains to be seen if there are other parameters that can be measured robustly, and whether other techniques (e.g., gravitational waves, x-ray reflection spectroscopy) can succeed where BH shadows fail. We leave these analyses for the future.


{\it Acknowledgments} $-$ 
We thank Sebastian Voelkel for fruitful discussions during the preparation of this manuscript. S.N. acknowledges support from the Alexander von Humboldt Foundation.


\bibliography{references}

\end{document}